# Propagation of transition fronts in nonlinear chains with non-degenerate on-site potentials

I.B. Shiroky and O.V. Gendelman*

Faculty of Mechanical Engineering, Technion, Haifa 32000, Israel

* - contacting author, ovgend@tx.technion.ac.il

We address the problem of a front propagation in chains with a bi-stable nondegenerate on-site potential and a nonlinear gradient coupling. For a generic nonlinear coupling, one encounters a special regime of transitions, characterized by extremely narrow fronts, far supersonic velocities of propagation and long waves in the oscillatory tail. This regime can be qualitatively associated with a shock wave. The front propagation can be described with the help of a simple reduced-order model; the latter delivers a kinetic law, which is almost not sensitive to fine details of the on-site potential. Besides, it is possible to predict all main characteristics of the transition front, including its shape and frequency and amplitude of the oscillatory tail. The numerical results are in a good agreement with the analytical predictions. The suggested approach allows one to consider the effects of an external pre-load and on-site damping. When the damping is moderate, the analysis remains in the frame of the reduced-order model. It is possible to consider the solution for the front propagating in the damped chain as a perturbation of the undamped dynamics. This approach yield reasonable predictions. When the damping is high, the transition front enters a completely different asymptotic regime. The gradient nonlinearity generically turns negligible, and the propagating front converges to the exact solution obtained from a simple linear continuous model.

Keywords: Front propagation, kinks, shock waves, Frenkel-Kontorova model, damping.

1. Introduction

Transition fronts, also known as phase boundaries, are common in systems in which the potential energy has more than one stable equilibrium. A broad variety of processes in actual systems and materials can be described by such switching of states. Among many possible applications, one finds dislocations in metals [1, 2, 3], dry friction [4], dynamics of carbon nano tubes foams [5], pulse propagation in cardio physiology [6], lattice distortions around twin boundaries [7], domain walls in ferro-electrics [8], crack propagation [9, 10], motion of fronts in semiconductor superlattices [11], surface reconstruction phenomena [12], calcium release in cells [13] and statistical mechanics [14].

Continuous models of the aforementioned systems with degenerate equilibria and without damping are often characterized by continuous spectrum of possible front velocities. In discrete case, the radiative damping should be balanced by the energy influx. Thus, a "kinetic relation" is established [15]. In the current paper we primarily deal with the discrete models and restrict ourselves by the simplest case of bi-stable model potentials. For certain processes, the bi-stability characterizes an on-site potential [1, 15, 16], while in others it acts between particles (gradient bi-stable potential) [17, 18]. Some systems can be described as energy conserving and therefore Hamiltonian models are considered [19, 1, 20], while in others dissipation is included [16, 21, 22, 23]. In the same time, the main reason for the ongoing growth in studies of bi-stable systems is the numerous possible configurations of the potentials in the system, with the focus on the variations in the shape of the bi-stable potential. The pioneering work of Frenkel and Kontorova [20] considered a sinusoidal on-site potential, thus introducing the discrete version of sine-Gordon equation. Smooth layouts of the on-site potentials (also referred to as fully nonlinear) are studied numerically in [24] and [25] and show that these might result in slower velocity of the defect propagation. In [16] a linear chain with the smooth on-site potential is studied by means of an approximate model. However, probably the most widely used on-site potential was introduced in the renowned work of Atkinson and Cabrera [1] where they replaced the nonlinear on-site potential with a piecewise parabolic potential, ending with a set of linear equations. The same choice was later made in the early works of Ishioka [22] and Celli and Flytzanis [26]. The case where both wells have the same curvatures has a well-known solution which is available through a direct Fourier transform, which is presented in [1] and further explored in [16, 15, 27]. The case where the curvatures are different requires an application of the Wiener-Hopf method as shown in [28]. Paper [18] studies a modification of the bi-parabolic potential by an inclusion of a non-convex region (a spinodal region) that smoothens the cusp of the pure bi-parabolic potential.

Dynamics of the transition fronts in the chains with the bistable on-site potential and nonlinear gradient coupling is much less explored. In weakly nonlinear lattices, directional waveguiding has been achieved by using cubic Kerr nonlinearities in nonhomogeneous systems [8,9]. In strongly nonlinear chains of elastically coupled rotational pendula, a steady front propagation was observed in [10]. Recent studies [29, 21] have addressed the case of a generic coupling with on-site dissipation and suggested a law that connects the transported energy with the velocity and dissipation ratio. The results were verified experimentally. A numerical study of a one-dimensional chain with a nonlinear coupling was presented in [30]. It was shown that even at relatively weak gradient nonlinearities the velocity of the kink propagation increases dramatically. It was also revealed numerically that for high values of nonlinear coupling $\beta$, the front velocity is proportional to $\sqrt{\beta}$.

In several works, dynamics of Frenkel-Kontorova-based systems is analyzed by means of equivalent reduced models. In [16] an approach named "active point theory" is applied to construct approximate solutions for the case of a damped chain with an on-site potential with cubic nonlinearity. A similar method is used in [31]. In [23] the "local mode approximation" is employed to reduce the damped Frenkel-Kontorova problem to a pendulum equivalent model. Generally, these works demonstrate a good agreement of the simplistic models with the numerical integration of the full multi-particle nonlinear equations.

In [32] the regime of front propagation dominated by the inter-particle cubic nonlinearity is studied analytically and verified numerically. It is figured out that the characteristics of the response change dramatically compared to the case of linear coupling: the front becomes very narrow with an extreme energy concentration, the front velocity approaches far supersonic velocities and the wavenumbers of tail oscillations are very low. These properties allow derivation of a rather simple yet accurate "kinetic relation" in which the specific details of the on-site potential structure don't play a significant role. Moreover, it is demonstrated that chains with different on-site potentials but with same general characteristics demonstrate similar velocity of defect propagation.

Current paper considers the aforementioned fast and narrow transition fronts in models with generic gradient nonlinear coupling. Analytical description is performed by means of an appropriately modified reduced-order model. We reveal possible effects of the chain pre-load on the transition velocity. Besides, modifications of the front dynamics due to inclusion of the on-site linear damping are described and explored analytically. The structure of the paper is as follows. In Section 2, the general methodology for the case of a generic nonlinear coupling is developed. The claim that the formulation is generic is verified through examples and the robustness with respect to variations in the on-site potential is examined. The model is also extended to consider the next-nearest-neighbor interactions. In Section 3, the problem of a pre-loaded chain is addressed. This problem is treated with the help of similar techniques as in the free chain, since the effect of preload can be considered as a modification of the coupling potential. It is demonstrated that the external pre-load applied to the chain can considerably modify the velocity of the front propagation. In Section 4, the on-site damping is considered. It is shown that the small damping can be considered as a perturbation of the conservative case. In the opposite limit of high damping, the front propagation can be described means of a simple continuous linear model.

## 2 Conservative bi-stable chains.
### 2.1 General treatment

We consider a chain with a bi-stable nondegenerate on-site potential [32, 19] with a generic nonlinear gradient coupling. The particular case of a cubic gradient nonlinearity was considered in [32]. Hamiltonian of this chain is written as follows:

$$H = \sum_{n=1}^{\infty}\left[\frac{p_n^2}{2} + U_1\left(\varphi_{n+1} - \varphi_n\right) + U_2\left(\varphi_n\right)\right]. \tag{1}$$

Here $\varphi_n$ is the displacement of the $n^{th}$ particle from the initial equilibrium state (meta-stable), $U_1(\varphi_{n+1} - \varphi_n)$ is the gradient potential of the interparticle interaction, $U_2(\varphi_n)$ is a non-degenerate bi-stable on-site potential; $p_n = \dot{\varphi}_n$, masses of all particles are set to unity. $U_2(\varphi_n)$ is defined by three main characteristics: the energetic effect $Q$, the height of the potential barrier $B$, and the coordinate difference between the stable and meta-stable states $\varphi^*$. Also, the minimum of the meta-stable state is set to $\varphi = 0$ without affecting the generality. Obviously, infinite number of possible bi-stable potentials have such characteristics, and we restrict ourselves to three typical shapes: piecewise parabolic, 4$^{th}$ order polynomial and 6$^{th}$ order polynomial (Figure 1). The details on these model on-site potentials are presented in Appendix.

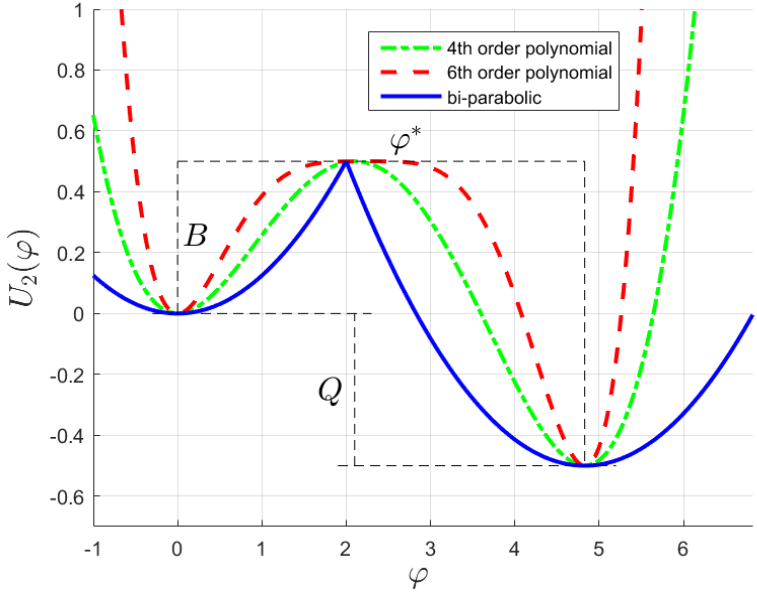

*Figure 1 - On-site nondegenerate potential $U_2(\varphi)$. Three possible approximations with the same basic shape parameters are presented: solid-blue – bi-parabolic potential, line-dotted green – 4$^{th}$ order polynomial, dashed red – 6$^{th}$ order polynomial.*

The gradient potential $U_1(\varphi_{n+1} - \varphi_n)$ can be a linear potential, as previously studied in several works [19, 15, 28]. However, if the gradient potential is nonlinear, we reveal numerically that the front propagation can enter a regime completely dominated by the coupling nonlinearity, rather than by the specific details of $U_2(\varphi_n)$. The example below involves the Lennard-Jones (LJ) coupling potential, $U_1(r) = \varepsilon\left[\sigma^{12}\left(r + 2^{1/6}\sigma\right)^{-12} - \sigma^6\left(r + 2^{1/6}\sigma\right)^{-6}\right]$. A typical response of the chain in presented in $n - \varphi$ plane for a fixed time instance (Figure 2). The only nonzero initial condition is the velocity of particle #1 – $\dot{\varphi}(0) = 10$. From here on, this condition is denoted in figures as "impulse 10".

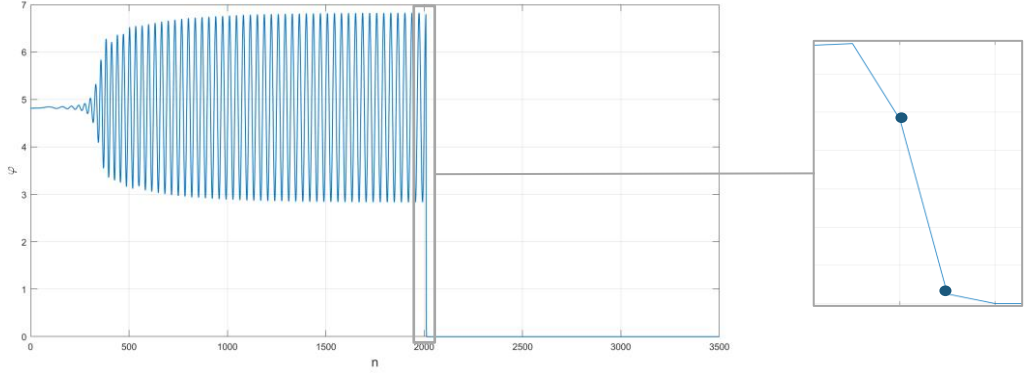

*Figure 2 - Dynamic response of the chain with LJ coupling, $\sigma = 20$, $\varepsilon = 2^{1/3}\sigma^2/18$, for $t = 700$. On-site potential: bi-parabolic with parameters: $Q = B = 0.5$, $\omega_0 = 0.5$, initial conditions: impulse 10. The inset depicts the front zone in details.*

First, we observe that the velocity is far supersonic. For example, for the response in Figure 2, the front propagates at a rate of 3.02. In contrast, for the same parameters of the on-site potential, but with a quadratic gradient potential, the front would propagate with a front velocity of 0.86. Further details on front velocities obtained with nonlinear gradient potentials are developed in subsections 2.2-2.3 below. Then, the transition area is extremely narrow and comprises only 1-2 particles. In Figure 2 a typical example with 2 particles is shown. Third, the oscillatory tail has a very large wavelength when compared to the narrow front area. In the discussed example, each period of oscillation within the tail consists of about 36 particles. The evidence for the dominance of the coupling potential in the front zone is obtained from the strain energy distribution along the chain. The expression for $\overline{e}_{U_1}$, the average distributions of the strain energy, is presented as follows:

$$\overline{e}_{U_1}(n) = \frac{1}{\tau} \int_{t_1}^{t_1+\tau} U_1\big(\varphi_n(t) - \varphi_{n-1}(t)\big) dt \qquad (2)$$

Here $\tau = 1/V$ is the characteristic time of transition, such that for particles within the transition region $\varphi_n(t) = \varphi_{n+1}(t+\tau)$. Typical numerical results for the chain with LJ coupling are presented in Figure 3. One can observe that the concentration is extremely high in the narrow transition area compared to the rest of the chain.

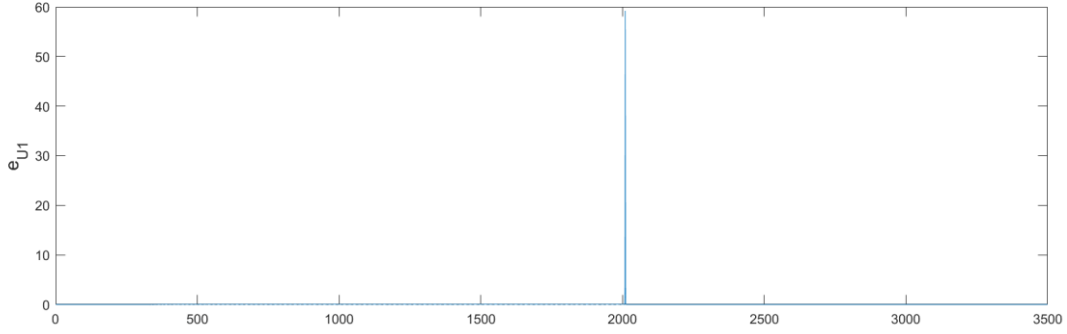

*Figure 3 – Nearest-neighbor interaction energy of the chain with LJ coupling term $\sigma = 20$, $\varepsilon = 2^{1/3}\sigma^2/18$ at $t = 700$; On-site potential: bi-parabolic with: $Q = B = 0.5, \omega_0 = 0.5$, initial conditions: impulse 10*

To construct the simplified model, we first assume the dominance of the nonlinear term (Figure 3) in the transition area, compared to contribution of the on-site potential (Figure 4). The maximal energy of the on-site potential is 0.5, which is less than 1% of the gradient potential. Therefore, we neglect $U_2(\varphi_n)$ from equation (1). It will be demonstrated that the on-site potential affects only the boundary conditions of the solution. Hence, we obtain the following Hamiltonian for the transitional area:

$$H \approx \sum_j \left[ \frac{\dot{\varphi}_j^2}{2} + U_1\left(\varphi_{j+1} - \varphi_j\right) \right] \quad (3)$$

Here $j$ are the indices of particles that belong to the transitional area.

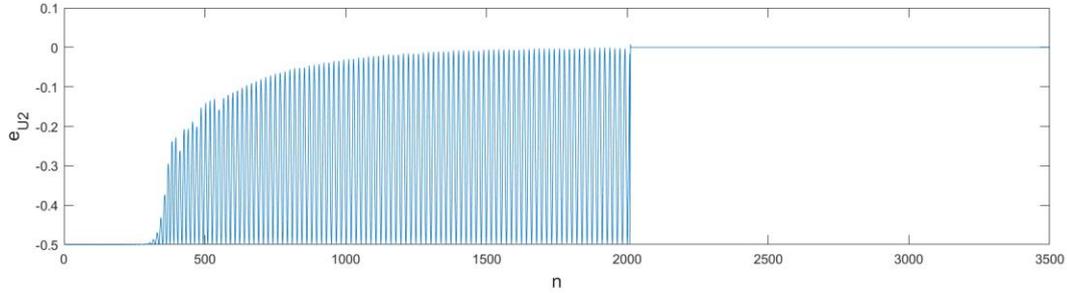

*Figure 4 – On-site interaction energy of the chain with LJ coupling term $\sigma = 20$, $\varepsilon = 2^{1/3}\sigma^2/18$ at $t = 700$; On-site potential: bi-parabolic with: $Q = B = 0.5, \omega_0 = 0.5$, initial conditions: impulse 10*

Then, as was mentioned earlier, the transition area is extremely narrow (see Figure 2). Therefore, only these few particles contribute in the summation in Eq. (3). We simplify the problem further and admit that, at large enough velocities, an effective description may be obtained by considering the rapid jump of a single particle from the meta-stable state to the vicinity of the stable state.

To complete the construction of the approximate model, we use the observation that the gradient in the transitional region is extremely steep when compared to the layout within the two wells in its vicinity. At its one edge the transitional region is attached to a nearly fixed particle that still has not left the metastable position $\varphi = 0$. At the other edge, the transition region is attached to the oscillatory tail. The energy density in the oscillatory tail is relatively low, and therefore its dynamics can be described in the framework of the linear dispersion relation (cf. [32]) presented in Figure 5. In a steady state propagation, the phase velocity ($V_{ph} = \omega/k$) must be equal to the front velocity ($V$) and when it is very high, the wavenumber $k$ in the tail is close to the left bandgap of the dispersion relation (Figure 5) and thus is very small. This comes in a considerable contrast to the extreme concentration of energy in the transition area, so one can admit that from the perspective of the transiting particle it is attached to an immobile point with coordinate $\Delta$, defined as the first maximum of the oscillatory tail behind the transition front (see Figure 6).

To find the value of $\Delta$ one should notice that close to the left bandgap of the dispersion relation, the group velocity is small. So, the energy transport through the oscillatory tail can be neglected and the energy released due to the front propagation is almost not transferred towards or from the front. The energy balance for an arbitrary particle $n$ in the oscillatory tail can be simply expressed as:

$$\frac{\dot{\varphi}_n^2}{2} + U_2(\varphi_n) = 0 \tag{4}$$

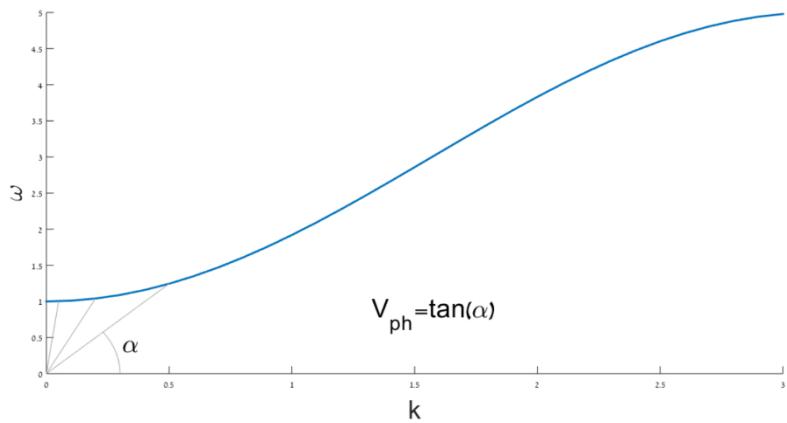

*Figure 5- A typical dispersion relation plot*

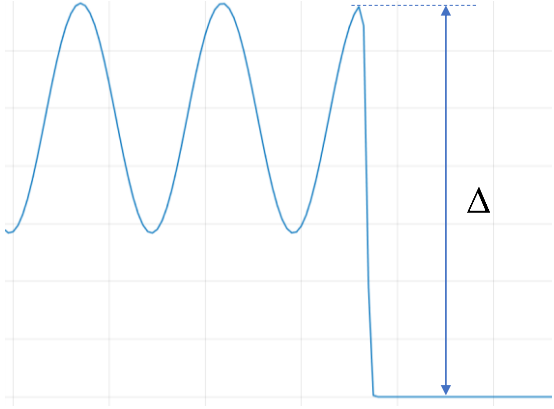

*Figure 6 - Definition of $\Delta$*

The peak of the oscillation, $\varphi_n = \Delta$, occurs when $\dot{\varphi}_n = 0$. Therefore, one has to extract $\Delta$ from the following relation:

$$U_2(\Delta) = 0 \qquad (5)$$

The resulting approximate single-DOF Hamiltonian for the particle inside the transition front is written as follows:

$$H = \frac{\dot{\varphi}^2}{2} + U_1(0-\varphi) + U_1(\varphi-\Delta) \qquad (6)$$

From Hamiltonian (6), one obtains:

$$dt = \frac{d\varphi}{\sqrt{2}\sqrt{U_1(0)+U_1(-\Delta)-U_1(-\varphi)-U_1(\varphi-\Delta)}} \qquad (7)$$

By using $V = 1/\int_{t=0}^{t(\varphi=\Delta)} dt$, the corresponding velocity is found from the following expression:

$$V \approx \frac{1}{\int_0^{\Delta} \frac{d\varphi}{\sqrt{2}\sqrt{U_1(0)+U_1(-\Delta)-U_1(-\varphi)-U_1(\varphi-\Delta)}}} \qquad (8)$$

Expression (8) is the general formulation of the approximate model for the case of the nonlinear gradient potential $U_1$. In subsections 2.2-2.3 we verify the result for two specific examples and test it for robustness to variations in the shape of the on-site potential.

One should notice, that although the treatment is general, the specific parameters for which the nonlinear regime is fully established vary with the parameters of the potentials. As a rule of thumb, this treatment holds for responses with front velocity $V > 2$.

### 2.2 Polynomial coupling potential in the form of Fermi-Pasta-Ulam (FPU).

After providing a general framework for analysis of systems with arbitrary nonlinear gradient potential $U_1$ and the nondegenerate bi-stable potential $U_2$, we address several specific cases. The first one is the gradient potential that is a generalized case of the cubic coupling studied in [32]. Here we propose a coupling potential that comprises quadratic, cubic and quartic terms. This potential can be seen as a Taylor expansion of more general potentials, and is presented in the following form:

$$U_1(r) = \frac{1}{2}r^2 + \frac{\alpha}{3}r^3 + \frac{\beta}{4}r^4 \tag{9}$$

$\alpha$ and $\beta$ are the stiffnesses of the quadratic and cubic nearest-neighbor springs respectively. Without loss of generality the linear stiffness is set to unity. In the regime dominated by the nonlinearities of the inter-particle interaction, the energy that is contained within the linear portion of coupling is negligible when compared to the nonlinear terms ($\alpha, \beta$). We neglect the linear coupling term, and by substitution of (9) into (6) we achieve the following Hamiltonian that describes the single particle dynamics in the region of the front:

$$H = \frac{\dot{\varphi}^2}{2} + \frac{\beta}{4}\left[\varphi^4 + (\varphi - \Delta)^4\right] + \frac{\alpha}{3}\left[-\varphi^3 + (\varphi - \Delta)^3\right] \tag{10}$$

By substituting $z = \varphi/\Delta$, integrating over the entire motion range $0 < z < 1$, the following solution is obtained for the SDOF model:

$$V = \frac{\Delta\sqrt{\beta}}{\sqrt{2}} \frac{1}{\int_0^1 \frac{dz}{\sqrt{\left[1 - z^4 - (1-z)^4\right] - \frac{4}{3}\frac{\alpha}{\beta\Delta}\left[1 - z^3 - (1-z)^3\right]}}} \tag{11}$$

One can integrate the expression (11) numerically for different sets of $\alpha, \beta$. However, a simple approximated expression can be obtained by defining a small parameter $\mu \equiv \frac{4}{3}\frac{\alpha}{\beta\Delta} \ll 1$:

$$V = \frac{\Delta\sqrt{\beta}}{\sqrt{2}} \frac{1}{\int_0^1 \frac{dz}{\sqrt{1 - z^4 - (1-z)^4}} - \frac{\mu}{2}\int_0^1 \frac{\left[-1 + z^3 + (1-z)^3\right]dz}{\left[1 - z^4 - (1-z)^4\right]^{\frac{3}{2}}} + O(\mu^2)} \tag{12}$$

Neglecting $O(\mu^2)$ terms and integration yield:

$$V \approx \frac{\Delta\sqrt{\beta}}{\sqrt{2}} \frac{1}{\left(K\left(\frac{\sqrt{2}}{4}\right) + \mu\frac{3}{7}E\left(\frac{\sqrt{2}}{4}\right)\right)} \approx f_1(\beta,\Delta) + f_2(\alpha,\beta)$$

$$f_1(\beta,\Delta) = \frac{1}{\sqrt{2}K\left(\frac{\sqrt{2}}{4}\right)}\Delta\sqrt{\beta}, \quad f_2(\alpha,\beta) = -\frac{4E\left(\frac{\sqrt{2}}{4}\right)}{7\sqrt{2}K\left(\frac{\sqrt{2}}{4}\right)^2}\frac{\alpha}{\sqrt{\beta}}$$

(13)

Here K and E are complete elliptic integrals of the first and the second kind respectively. This analysis leads to a conclusion that the $\alpha-\beta$ problem can be treated as two separate problems; The basic problem is the pure $\beta$ problem ($f_1$) which determines the nominal velocity through values of $\beta$ and $\Delta$. The $\beta$ contribution can be effectively described by the following scaling law:

$$V(\alpha=0) = f_1(\beta,\Delta) \sim \Delta\sqrt{\beta} \tag{14}$$

The second problem is the contribution of $\alpha$ to the nominal velocity - $f_2$. The contribution can be described as a modification of the velocity established and dominated by $\beta$ term alone according to the following law:

$$V - V(\alpha=0) = f_2(\alpha,\beta) \sim -\frac{\alpha}{\sqrt{\beta}} \tag{15}$$

It was found from numerical simulations that in order to obtain a very good quantitative agreement, it is enough to multiply $f_1$ and $f_2$ by constant coefficients $\gamma_1 \approx 1.33$ and $\gamma_2 \approx 1.6$. Necessity of these corrections stems from the simplification and assumptions taken. These factors were extracted from numerous numerical simulations for varying $\alpha$, $\beta$, $\Delta$ and although their chosen values should be treated as an assumption, they remained nearly constant for the range of the front velocities $V > 2$. Therefore, it is reasonable to assume that, with good accuracy, these factors are nearly constant.

The basic problem of pure cubic nonlinearity without an $\alpha$ term (14) was examined in [32], so here we explore the effect of the $\alpha$ term (15). In Figure 7 the modification of the front velocity is presented for different $\beta$ and is compared to the analytical predictions. It is seen that for $\beta \geq 0.2$ the approximation is within a reasonable tolerance from the numerical results. At $\beta = 0.1$ the approximation fails to describe the front velocity accurately for positive $\alpha$. This combination of parameters corresponds to low front velocities, that lie beyond the scope of validity of the simplified model.

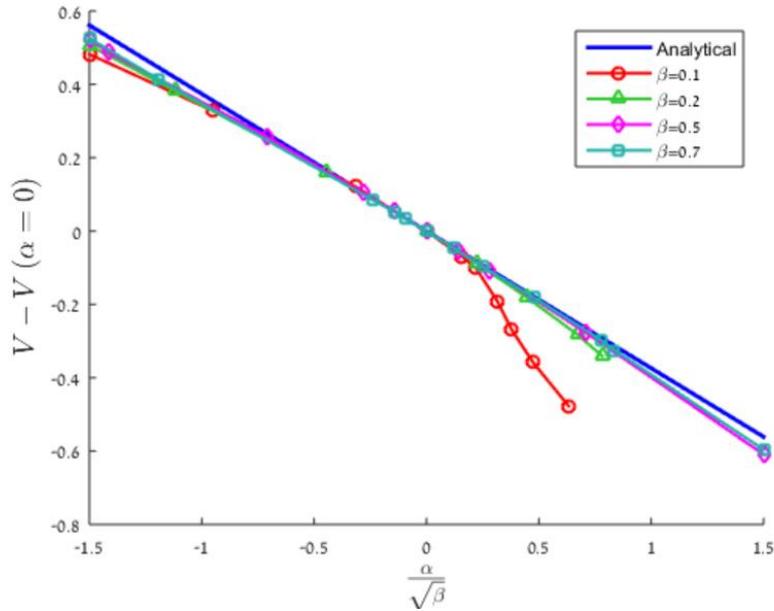

*Figure 7 – Change in front velocity due to $\alpha$; solid blue – analytical estimation, 'o' red - $\beta = 0.1$, triangles green - $\beta = 0.2$, diamonds purple - $\beta = 0.5$, squares turquoise - $\beta = 0.7$; On-site potential: bi-parabolic with parameters: $Q = 0.5$, $B = 0.5$, $\omega_0 = 0.5$*

Quite remarkably, in expression (13), the contribution of $\alpha$ to the front velocity doesn't involve any characteristics of the on-site potential. Unlike the $\beta$ contribution that depends on knowing the potential shape (through expression (5)), the $\alpha$ contribution can be determined for all potentials. To examine the legitimacy of this statement, we present the results for three bi-stable on-site potentials defined above (Figure 1) in Figure 8. All results collapse on the same line with a considerable accuracy.

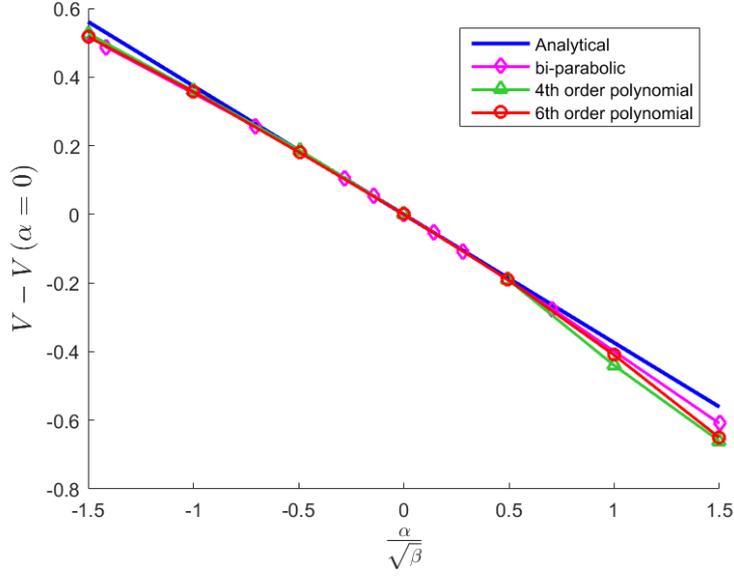

*Figure 8 – Change in the front velocity versus $\alpha$; solid blue – analytical estimation, diamonds purple – bi-parabolic potential; triangles green – 4$^{th}$ order polynomial potential; circles red – 6$^{th}$ order polynomial potential, common parameters: $Q = 0.5$, $B = 0.5$, $\varphi^* = 4.82$*

### 2.3 Lennard-Jones potential

Realistic coupling potentials are more complex than polynomials. Here we address the Lennard Jones (LJ) potential that is one of the most common models for actual inter-particle reactions. Let,

$$U_1(r) = U_{LJ}(r) = \varepsilon\left[\left(\frac{\sigma}{r+r^*}\right)^{12} - \left(\frac{\sigma}{r+r^*}\right)^{6}\right] \quad (16)$$

To set the force to zero at $r = 0$, the following condition should be satisfied:

$$\frac{dU_{LJ}}{dr}(r=0) = 0 \rightarrow r^* = 2^{1/6}\sigma \quad (17)$$

A typical response of a chain with LJ coupling was presented in Figure 2. Qualitatively, the nature of the response resembles the strongly nonlinear regime with a cubic nonlinearity, i.e. the narrow transition front with extremely high energy concentration, low wavenumbers in the oscillatory tail, and far supersonic responses. Thus, one can expect that the analytical approach that was used in the case of $\alpha - \beta$ nonlinearity is applicable in this case as well.

Here we employ the SDOF approximation (5), (8) to study the asymptotic behavior of the chain with the LJ potential. This approximation is expected to be reasonable at high velocities where the assumption that a single particle participates in the transition process simultaneously is close to reality. Let us consider the following SDOF Hamiltonian:

$$H = \frac{\dot{\varphi}^2}{2} + U_{LJ}(-\varphi) + U_{LJ}(\varphi - \Delta) \quad (18)$$

One obtains the following expression for the front velocity:

$$V \approx \frac{1}{\int_0^\Delta \frac{d\varphi}{\sqrt{2}\sqrt{U_{LJ}(0)+U_{LJ}(-\Delta)-U_{LJ}(-\varphi)-U_{LJ}(\varphi-\Delta)}}} \tag{19}$$

The argument of the square root in (19) can be expanded to a Taylor series in the following way:

$$U_{LJ}(0)+U_{LJ}(-\Delta)-U_{LJ}(-\varphi)-U_{LJ}(\varphi-\Delta) \approx$$
$$\varepsilon \left[ \frac{6\sigma^6}{\left(\Delta-2^{1/6}\sigma\right)^7} - \frac{12\sigma^{12}}{\left(\Delta-2^{1/6}\sigma\right)^{13}} \right] \varphi \tag{20}$$

The substitution of (20) into (19) yields:

$$V \approx \frac{1}{2^{17/12}\sqrt{3}\sqrt{\Delta}} \sqrt{\frac{\sigma^{14}}{\left(\sigma-2^{-1/6}\Delta\right)^{13}} - \frac{\sigma^8}{\left(\sigma-2^{-1/6}\Delta\right)^7}} \tag{21}$$

We deduce from (21) that the velocity tends to infinity for $\sigma \to 2^{-1/6}\Delta$. Also, we see that no simple scaling law exists between the velocity and governing parameters $\sigma, \Delta$.

To study the behavior for relatively low velocities, we adopt the previously studied $\alpha-\beta$ model as an approximation for the LJ potential, by means of a Taylor expansion of the full potential. Taylor expansion of (16) yields:

$$U = U(0) + \frac{1}{2}\frac{18\varepsilon}{2^{1/3}\sigma^2}r^2 - \frac{1}{3}\frac{189\varepsilon}{2^{1/2}\sigma^3}r^3 + \frac{1}{4}\frac{1113\varepsilon}{2^{2/3}\sigma^4}r^4 + O(r^5) \tag{22}$$

Without affecting the generality, we set the coefficient of linear coupling to 1, achieve a constraint on $\varepsilon$ and express $\alpha, \beta$:

$$\left. \frac{18\varepsilon}{2^{1/3}\sigma^2} = 1 \right| \to \varepsilon = \frac{2^{1/3}\sigma^2}{18} \to \alpha = -\frac{21}{2^{7/6}\sigma} \qquad \beta = \frac{371}{6 \cdot 2^{1/3}\sigma^2} \tag{23}$$

This leads to the approximated $\alpha-\beta$ potential that is a function of a single parameter $\sigma$:

$$U_{\alpha-\beta} = U(0) + \frac{1}{2}r^2 + \frac{1}{3}\frac{-21}{2^{7/6}\sigma}r^3 + \frac{1}{4}\frac{371}{6 \cdot 2^{1/3}\sigma^2}r^4 \tag{24}$$

The $\alpha-\beta$ approximation is expected to work at low $r$, where the LJ potential is probed in a narrow envelope around $r=0$. This correlates with high $\sigma$. In turn, high $\sigma$ yields small values of $\alpha-\beta$. This leads to a conclusion that at high $\sigma$ the value of velocity with LJ potential asymptotically converges to the velocity with a linear potential $1/2\,r^2$. This value has a closed form analytical expression when the on-site potential is piecewise parabolic that was derived in [15, 31] and rescaled to the parametrization of the current analysis in [32]. In Figure 9 the results of the velocity for the chain with LJ potential are presented. Indeed, it is seen that the full solution converges to the $\alpha-\beta$ model at low velocities and to the SDOF model at high velocities.

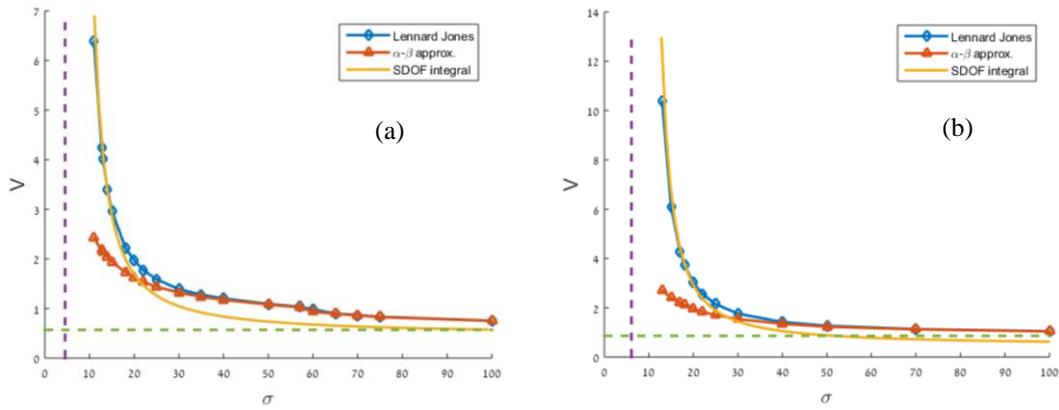

*Figure 9 - Front velocity of a chain with LJ interaction; blue-diamonds – numerical result of the full LJ potential with $\sigma = 20$, red-triangles – numerical result of the approximated $\alpha - \beta$ potential, solid yellow – SDOF model solution, horizontal dashed green – asymptotic value of the linearly coupled chain, vertical dashed purple – asymptotic $\sigma$ for which $V \to \infty$. On-site potential - bi-parabolic with $B = 0.5$, $\omega_0 = 0.5$; (a) $Q = 0.1$, (b) $Q = 0.5$.*

A main assumption that is taken during the analysis is that the only parameter of the on-site potential that has a direct effect on the front velocity is $\Delta$ (Figure 6). In the current case, the relationship can't be scaled by a simple law as can be seen even from approximation (21). Hence, we check the self-consistency of the assumption by numerically extracting the relationship between $V$ and $\Delta$ for different on-site potentials (Figure 10), and discover that all results coincide accurately on the same curve.

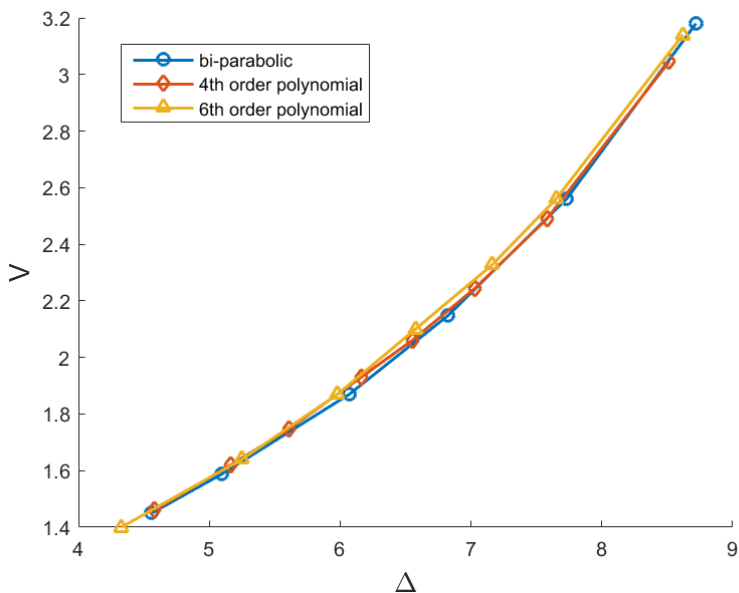

*Figure 10 - Front velocity of a chain with LJ interaction for different on-site potentials; 'o' – bi parabolic potential, diamonds – 4th order polynomial potential, triangles – 6th order polynomial potential; common parameters: $\sigma = 25$, $B = 0.5$*

It was seen that transitions in chains with a LJ nearest neighbor bonds result in steep increase of front velocity for low $\sigma$. One of the reasons that the actual response will probably not exhibit such an unbounded increase in front velocity is the presence of interactions with distant neighbors, rather than only a nearest-neighbor interaction. Let us examine a model with a second nearest neighbor LJ interaction as shown in Figure 11.

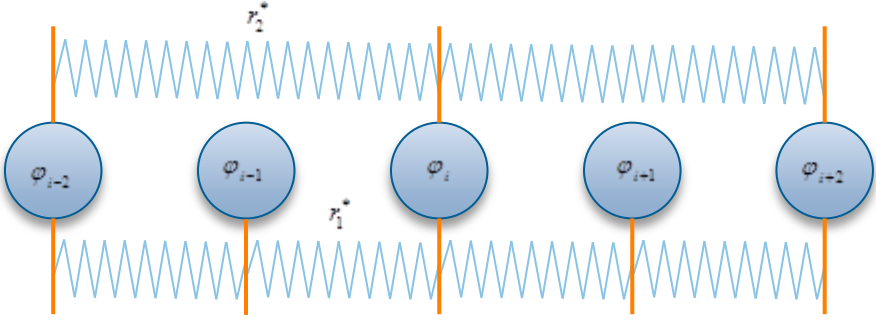

*Figure 11 - A model of chain with second nearest neighbor LJ interactions*

As a particular case, we set the equilibrium distance $r_2^*$ to $r_2^* = 2r_1^*$ (see Figure 11). Furthermore, the following relationship is assumed:

$$\sigma_2 = 2\sigma_1 \quad (25)$$

According to (22), the linear portion of the stiffness of a LJ interaction is:

$$k_m = c_m^2 = \frac{18\varepsilon_m}{2^{1/3}\sigma_m^2} \quad m=1,2 \quad (26)$$

The speed of sound of this lattice is governed by the effective linear stiffness:

$$c = \sqrt{k_1 + 4k_2} \quad (27)$$

We substitute the rescaling relationships (25), (26) into (27) and set the equivalent speed of sound to $c=1$, and obtain:

$$c = \sqrt{\frac{18}{2^{1/3}}\left(\frac{\varepsilon_1}{\sigma_1^2} + 4\frac{\varepsilon_2}{\sigma_2^2}\right)} = 1 \quad \rightarrow \quad \varepsilon_1 + \varepsilon_2 = \frac{2^{1/3}}{18}\sigma_1^2 \quad (28)$$

We introduce the ratio $\eta$ as the coupling strength ration of the two interactions ($\varepsilon_1, \varepsilon_2$) and obtain expressions for $\varepsilon_1, \varepsilon_2$:

$$\eta = \frac{\varepsilon_2}{\varepsilon_1} \quad \rightarrow \quad \varepsilon_1 = \frac{2^{1/3}}{18(1+\eta)}\sigma_1^2, \quad \varepsilon_2 = \frac{2^{1/3}\eta}{18(1+\eta)}\sigma_1^2 \quad (29)$$

In Figure 12 a typical influence of $\eta$ on the front velocity is presented numerically. It is seen that increase in the portion of the second nearest-neighbor interaction causes a decrease in the front velocity.

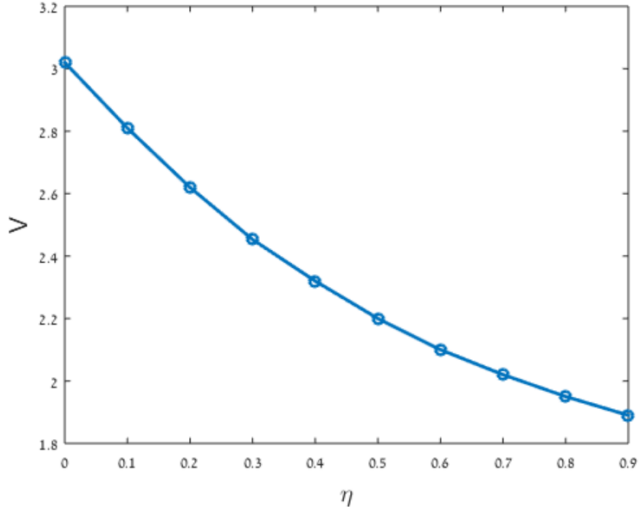

Figure 12 – $V = V(\eta)$; Front velocity of a chain with LJ interactions ($\sigma = 20$) with nearest and second nearest neighbors; On-site potential - bi-parabolic with $Q = 0.5, B = 0.5, \omega_0 = 0.5$.

## 3  The front propagation in the pre-loaded chain

In this section, we regard the front propagation in the chain under an external constant forcing. One can expect that a preload of a lattice, either tension or compression, might have an effect on the velocity of front propagation. It can have practical applications in controlling speed of reactions or by conducting experiments which estimate properties of a lattice. The pre-load effect is analyzed in two cases: the pure cubic coupling and the LJ coupling. We demonstrate that preload problems can be equivalently described and analyzed in the same manner as the basic Hamiltonian case (1).

### 3.1  Chain with pure cubic inter particle interaction

Let us address the case of the chain with an inter-particle interaction the consists of linear and cubic terms and a preload $f$ that is applied on the edges of the chain:

$$\begin{aligned} \ddot{\varphi}_i + (2\varphi_i - \varphi_{i+1} - \varphi_{i-1}) + \beta\left[(\varphi_i - \varphi_{i+1})^3 + (\varphi_i - \varphi_{i-1})^3\right] &= F(\varphi_i) & i \neq 0, n \\ \ddot{\varphi}_0 + (\varphi_0 - \varphi_1) + \beta(\varphi_0 - \varphi_1)^3 &= F(\varphi_0) - f & i = 0 \\ \ddot{\varphi}_n + (\varphi_n - \varphi_{n-1}) + \beta(\varphi_n - \varphi_{n-1})^3 &= F(\varphi_n) + f & i = n \end{aligned} \quad (30)$$

In equilibrium, we denote:

$$\varphi_i^0 - \varphi_{i-1}^0 = \delta \quad (31)$$

By substitution of (31) into (30) for $i = 0, n$ the expression for $\delta$ as a function of $\beta, f$ is obtained:

$$\delta + \beta\delta^3 = f \quad (32)$$

The following coordinate transform is further employed:

$$\varphi_i = u_i + i\delta \tag{33}$$

Let us define the on-site force in the following way:

$$F(\varphi_i) = F(u_i + i\delta) = \tilde{F}(u_i) = \begin{cases} -\omega_0^2 u_i & u_i \leq b \\ -\omega_0^2(u_i - \varphi^*) & u_i > b \end{cases} \tag{34}$$

By substituting (33) and (34) into (30) for $i \neq 0, n$ we obtain:

$$\ddot{u}_i + (1 + 3\beta\delta^2)(2u_i - u_{i+1} - u_{i-1}) + 3\beta\delta\left[(u_i - u_{i-1})^2 - (u_i - u_{i+1})^2\right] + \\ + \beta\left[(u_i - u_{i-1})^3 + (u_i - u_{i+1})^3\right] = \tilde{F}(u_i) \tag{35}$$

By denoting $\tilde{c}^2 = 1 + 3\beta\delta^2$, $\tilde{\alpha} = 3\beta\delta$ one arrives to the following equations:

$$\ddot{u}_i + \tilde{c}^2(2u_i - u_{i+1} - u_{i-1}) + \\ \tilde{\alpha}\left[(u_i - u_{i-1})^2 - (u_i - u_{i+1})^2\right] + \\ \beta\left[(u_i - u_{i-1})^3 + (u_i - u_{i+1})^3\right] = \tilde{F}(u_i) \tag{36}$$

This re-parameterization leads to a conclusion that the problem with a preload force $f$, (30), is equivalent to the $c - \alpha - \beta$ problem, with $\tilde{c}, \tilde{\alpha}$ that satisfy:

$$\frac{1}{3\beta}\tilde{\alpha} + \frac{1}{27\beta^2}\tilde{\alpha}^3 = f \\ \tilde{c}^2 = 1 + \frac{\tilde{\alpha}^2}{3\beta} \tag{37}$$

$\tilde{\alpha}$ can be expressed explicitly:

$$\tilde{\alpha} = \frac{1}{2}\upsilon - \frac{6\beta}{\upsilon}, \quad \upsilon = \left(108f\beta^2 + 12\sqrt{81f^2\beta^4 + 12\beta^3}\right)^{\frac{1}{3}} \tag{38}$$

Therefore, the front velocity change due to an application of preload $f$ can be expressed in terms of $f, \beta$. The estimation of the front velocity is presented in (39). One admits that positive $f$ (tension) causes deceleration of the front propagation, whereas the compression results in an accelerated response.

$$V = V(f=0) - \\ -0.374\frac{1}{\sqrt{\beta}}\left(\frac{1}{2}\left(108f\beta^2 + 12\sqrt{81f^2\beta^4 + 12\beta^3}\right)^{\frac{1}{3}} - \frac{6\beta}{\left(108f\beta^2 + 12\sqrt{81f^2\beta^4 + 12\beta^3}\right)^{\frac{1}{3}}}\right) \tag{39}$$

In Figure 13 both analytical and numerical results for the front velocity versus $f$ are presented for different values of $\beta$. The match is very good for large $\beta$ (which correspond to high front velocities).

At low front velocities (low $\beta$ and high $f$), a discrepancy between numerical and analytical results is observed.

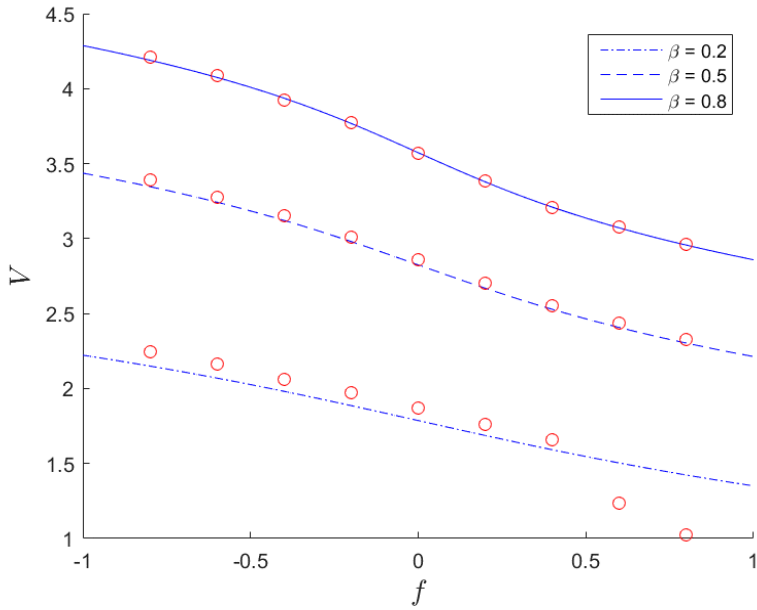

*Figure 13 – Front velocity as a function of $f$. Line dotted line - $\beta = 0.2$, dashed line - $\beta = 0.5$, solid line - $\beta = 0.8$; Bi-parabolic on-site potential with parameters: $Q = 0.5$, $B = 0.5$, $\omega_0 = 0.5$*

In Figure 14 we examine to robustness of the solution to a modification of the on-site potential shape. Due to the equivalence between $\alpha - \beta$ and preloaded $\beta$ models, one admits that the change of velocity due to preload is invariant to the shape of on-site potential. As it was shown in (39), the velocity change is only affected by the $\beta$ and $f$. This may give rise to experiments that quantify the strength of cubic $\beta$ nonlinearity by preloading a lattice, without knowing the actual structure of the on-site potential.

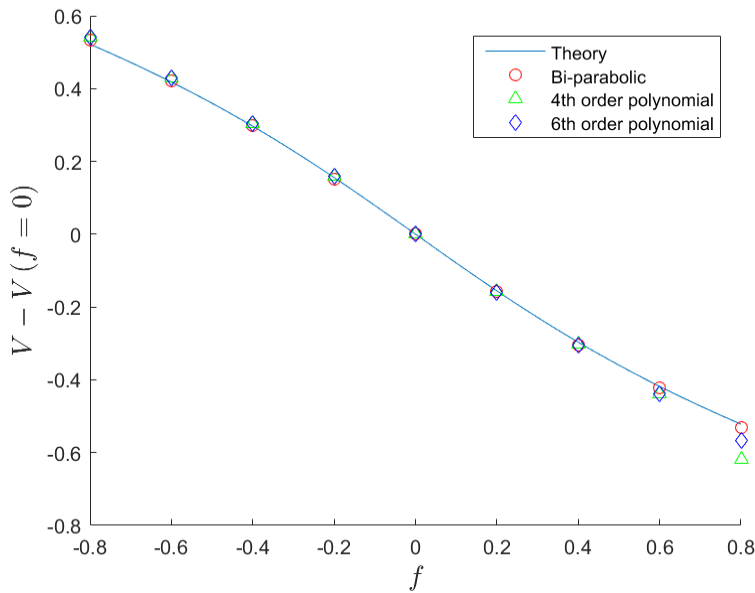

*Figure 14 – Change in front velocity due to preload $f$ for different potentials; solid blue – analytical estimation, 'o' red – bi-parabolic potential, triangles – 4th order polynomial potential, diamonds – 6th order potential. Common parameters: $\beta = 0.5$, $Q = 0.5$, $B = 0.5$, $\varphi^* = 4.82$*

### 3.2 The chain with LJ coupling under the pre-load.

We consider the model of chain with a LJ inter-particle interaction under pre-load $f$, that results in the following equations of motion:

$$\begin{aligned}
\ddot{\varphi}_i - F_{LJ}(\varphi_i - \varphi_{i-1}) + F_{LJ}(\varphi_{i+1} - \varphi_i) &= F(\varphi_i) \\
\ddot{\varphi}_0 + F_{LJ}(\varphi_1 - \varphi_0) &= F(\varphi_0) - f \\
\ddot{\varphi}_n - F_{LJ}(\varphi_n - \varphi_{n-1}) &= F(\varphi_i) + f
\end{aligned} \tag{40}$$

Here, $F_{LJ}(r) = -\dfrac{dU_{LJ}(r)}{dr}$

We again adopt (31), (33), (34) and achieve the following set of equations:

$$\begin{aligned}
\ddot{u}_i - F_{LJ}(u_i - u_{i-1} + \delta_{LJ}) + F_{LJ}(u_{i+1} - u_i + \delta_{LJ}) &= \tilde{F}(u_i) \\
\ddot{u}_0 + F_{LJ}(u_1 - u_0 + \delta_{LJ}) - F_{LJ}(\delta_{LJ}) &= \tilde{F}(u_0) \\
\ddot{u}_n - F_{LJ}(u_n - u_{n-1} + \delta_{LJ}) + F_{LJ}(\delta_{LJ}) &= \tilde{F}(u_n)
\end{aligned} \tag{41}$$

An equivalent potential $\hat{U}(r)$ that can provide equation (41) can be constructed to as follows:

$$\hat{U}_{LJ}(r) = \varepsilon\left[\frac{\sigma^{12}}{(r+\delta_{LJ}+r^*)^{12}} - \frac{\sigma^6}{(r+\delta_{LJ}+r^*)^6} + \left[\frac{12\sigma^{12}}{(\delta_{LJ}+r^*)^{13}} - \frac{6\sigma^6}{(\delta_{LJ}+r^*)^7}\right]r\right] \tag{42}$$

Now, the problem can be formulated in the form of (1) with $U_1 = \hat{U}_{LJ}$. Therefore, we can once again adopt a SDOF Hamiltonian of the following form:

$$H = \frac{\dot{\varphi}^2}{2} + \hat{U}_{LJ}(-\varphi) + \hat{U}_{LJ}(\varphi - \Delta) \tag{43}$$

One obtains the following expression for the front velocity:

$$V = \frac{1}{\displaystyle\int_0^\Delta \frac{d\varphi}{\sqrt{2}\sqrt{\hat{U}_{LJ}(0) + \hat{U}_{LJ}(-\Delta) - \hat{U}_{LJ}(-\varphi) - \hat{U}_{LJ}(\varphi-\Delta)}}} \tag{44}$$

Eq. (44) can be solved approximately by applying a Taylor expansion on the root argument, similarly to (20). The resulting solution is of the following form:

$$V \approx \frac{\sqrt{f + \dfrac{1}{3\cdot 2^{5/6}}\left[\dfrac{\sigma^{14}}{\left(\sigma - 2^{-1/6}(\Delta - \delta_{LJ})\right)^{13}} - \dfrac{\sigma^8}{\left(\sigma - 2^{-1/6}(\Delta - \delta_{LJ})\right)^7}\right]}}{2\sqrt{\Delta}} \tag{45}$$

From, (45) we understand the behavior of the velocity as a function of preload $f$. $\delta_{LJ}$ is a monotonous function of $f$. The velocity grows rapidly when the quantity $\sigma - 2^{-1/6}(\Delta - \delta_{LJ})$ decreases. Hence, a negative preload causes acceleration while a positive preload causes deceleration.

The effect of a preload force on a LJ chain is presented graphically in Figure 15. Three preload cases are shown: positive, negative and no preload. The front velocity is plotted as a function of $\Delta$. For each case the numerical results (markers) and the integration of the approximate SDOF model (44) (curves) are shown. Generally, the analytical curves are in a fair agreement with the numerically obtained values.

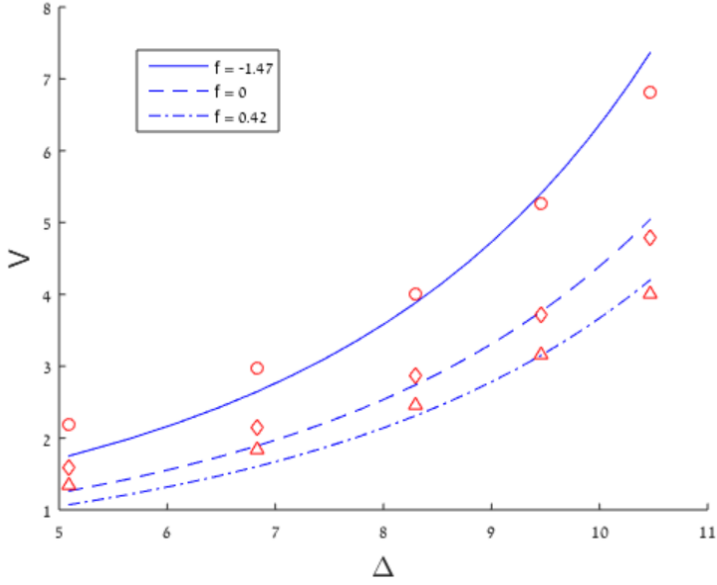

Figure 15 – $V = V(\Delta)$ of a chain with LJ interaction and preload for different values of preload; $f = -1.47$ : solid blue – SDOF model, 'o' – numerical; $f = 0$ : dashed blue – SDOF model, 'diamonds' – numerical; $f = 0.42$ : dash-dotted blue – SDOF model, 'triangles – numerical; Gradient potential: LJ with: $\sigma = 25$; numerical results were obtained for a bi-parabolic on site potential with $B = 0.5$, $\omega_0 = 0.5$ ( $Q$ varies to dictate $\Delta$ ).

## 4  Transition fronts in the chain with on-site linear damping

So far, we have treated conservative systems, in which the front propagates due to an internal energy transform. In real systems, in many cases, some sort of dissipative mechanism exists. Here we present an approach that allows an inclusion of an on-site linear damping in the previous discussion.

### 4.1  The case of low damping.

We assume that the only effect of the on-site damping on the front velocity is through a modification of $\Delta$ for the first particle within the stable well (Figure 16). Therefore, the rest of the energy is dissipated along the oscillatory tail has no impact on the velocity of propagation. The only contribution of the oscillatory tail in the simplified model is related to determination of parameter $\Delta$ in (8).

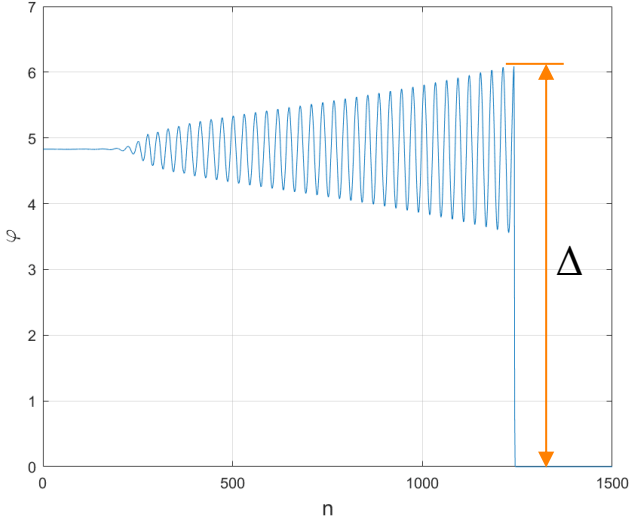

*Figure 16 – A dynamic response in the presence of on-site damping; Gradient potential: cubic with $\beta = 0.5$ ; On-site potential: bi-parabolic with $Q = 0.5, B = 0.5, \omega_0 = 0.5$ ; damping coefficient: $\xi = 0.005$, initial conditions: Impulse 10*

We assume that for low value of onsite damping coefficient $\xi$, the dynamics of the particle within the transitional region can be approximately described by the same single DOF Hamiltonian (6). During the motion of the particle, energy is dissipated from the system, and its total amount can be expressed as follows:

$$W_\xi = \xi \int_0^\Delta \dot{\varphi} d\varphi = \sqrt{2}\xi \int_0^\Delta \sqrt{U_1(0) + U_1(-\Delta) - U_1(-\varphi) - U_1(\varphi - \Delta)} d\varphi \tag{46}$$

The value $\Delta$ is found through a modification of (5). In the current case, the energy dissipated during the motion through the front has to be included in the energetic balance. So, the modified balance of energy for the first particle that has entered the stable well can be expressed as follows:

$$U_2(\Delta) + W_\xi = 0 \tag{47}$$

After $\Delta$ is extracted from equation (47), the front velocity can be determined from (8). This treatment is general for any combination of $U_1, U_2$. Here, we bring a specific example with the following selection of potentials: $U_1 = \frac{r^2}{2} + \frac{\beta r^4}{4}$ , $U_2$ a piecewise parabolic potential (A1). However, when the nonlinearity is dominant, and the damping is small enough, we can neglect the contribution of the quadratic term. The work done by the damping in this case is calculated as:

$$W_\xi = \frac{\xi \Delta^3 \sqrt{\beta}}{\sqrt{2}} \int_0^1 \sqrt{1 - z^4 - (z-1)^4} dz = \frac{I_1 \xi \Delta^3 \sqrt{\beta}}{\sqrt{2}}, \quad I_1 = \left(\frac{7}{3}\mathrm{K}\left(\frac{\sqrt{2}}{4}\right) - 2\mathrm{E}\left(\frac{\sqrt{2}}{4}\right)\right) \tag{48}$$

Substitution of equation (48) and of the expression for the stable branch of the bi-parabolic on-site potential (A1) into (47) yields:

$$\frac{\omega_0^2}{2}(\Delta - \varphi^*)^2 - Q + \frac{I_1}{\sqrt{2}c_d}\xi\Delta^3\beta = 0 \qquad (49)$$

The value of $\Delta$ can be extracted for any set of parameters. Here, unlike in the conservative case, its value depends not only on the parameters of potential $U_2$ alone, but also on $\beta$ and $\xi$. In (49) a correction factor $c_d$ is introduced, which is a result of approximations and assumptions that were taken in the estimation of $W_\xi$. Yet, for the selected potentials, its value was found numerically to be nearly constant $c_d \approx 1.5$, from verifications of the expression for different $\xi$ and $\beta$. Once the value of $\Delta$ is extracted, the velocity is found from the same expression as in the conservative case (8). The results for the test case of a bi-parabolic on-site potential and a cubic gradient potential are presented in Figure 17. The numerical results are in a good agreement with the analytical model for front velocities which are bigger than 2, which complies with all other findings in this work.

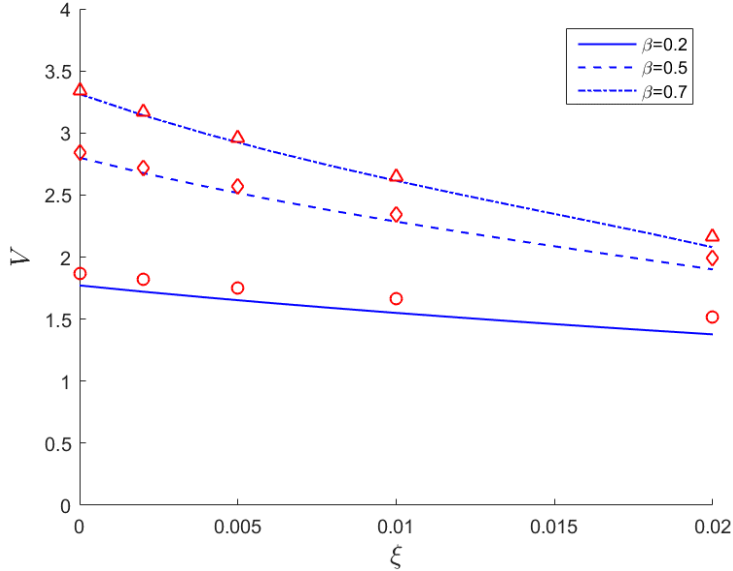

*Figure 17 – Front velocity in the presence of on-site damping; Solid line: analytical result for $\beta = 0.2$, dashed line: analytical result for $\beta = 0.5$, line-dotted line: analytical result for $\beta = 0.7$; gradient potential: cubic; on-site potential: bi-parabolic with $Q = 0.5, B = 0.5, \omega_0 = 0.5$.*

For $\xi \to 0$ the following approximation for $\Delta$ is obtained:

$$\Delta = \Delta_0 - \frac{(\omega_0\varphi^* + \sqrt{2Q})^3 I_1}{2c_d\sqrt{Q}\omega_0^4}\xi\sqrt{\beta}, \quad \Delta_0 = \varphi^* + \frac{\sqrt{2Q}}{\omega_0} \qquad (50)$$

Thus, the approximate expression for the change in velocity due to incorporation of on-site damping is:

$$V - V(\xi = 0) \approx -\frac{\gamma_1(\omega_0\varphi^* + \sqrt{2Q})^3 I_1 I_2}{2c_d\sqrt{Q}\omega_0^4}\xi\beta, \quad I_2 = \left(\sqrt{2}K\left(\frac{\sqrt{2}}{4}\right)\right)^{-1} \qquad (51)$$

It turns out that the modification of the front velocity due to inclusion of the on-site damping is proportional to $\beta$ and $\xi$ when the damping is small. To further justify the claim that the inclusion of damping is a perturbation of the Hamiltonian system, we examine the kinetic energy of the system. In Figure 18 the dependencies of total kinetic energy and the kinetic energy in the front region are shown. It is seen that the kinetic energy of the front slightly decreases as the damping increased and complies to the perturbative analytical result. On the other hand, the total kinetic energy is asymptotically infinite as the damping tends to zero, and can't be related to the perturbative nature of the damping effect on the dynamics. This is yet another evidence that the response in the strongly nonlinear regime is mostly affected by the internal energy conversion within the front region (the kink), rather than on the dynamics of the entire chain.

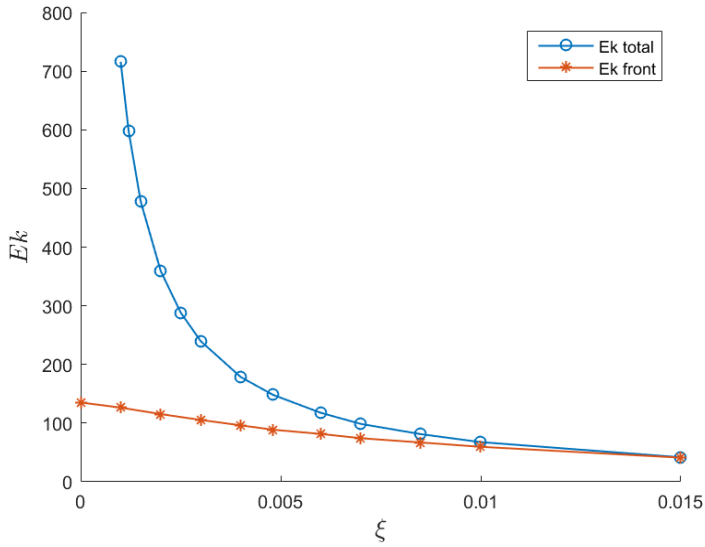

*Figure 18 – Kinetic energy as a function of damping coefficient, 'o'- total kinetic energy, '*' - kinetic energy of the front; gradient potential: cubic with $\beta = 0.5$; on-site potential: bi-parabolic with $Q = 0.5, B = 0.5, \omega_0 = 0.5$. Initial conditions: Impulse 10.*

### 4.2 The case of large damping

Here we address the opposite case of the large damping. A typical response is shown in Figure 19. We see that when damping is high enough, it completely suppresses the oscillations in the tail. Moreover, the front area becomes very wide (in this example – 20 particles). Therefore, it is reasonable to neglect the cubic coupling due to the low gradient and to adopt a continuum model to represent the dynamics in this case.

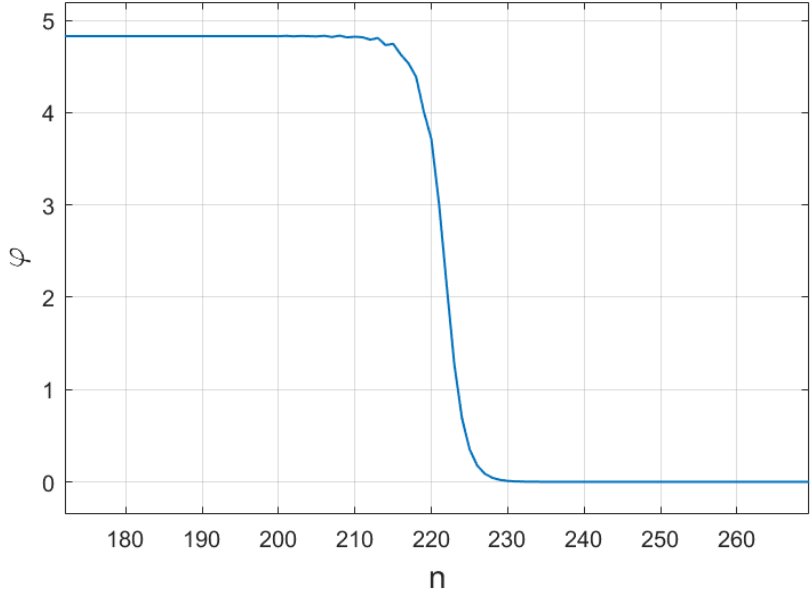

*Figure 19 – A typical response in the presence of high damping; gradient potential: cubic with $\beta = 0.5$; on-site potential: bi-parabolic with $Q = 0.5, B = 0.5, \omega_0 = 0.5$. Damping coefficient - $\xi = 0.3$; Initial conditions: Impulse 10.*

The kink is smooth enough to consider the continuum limit. Besides, all gradients are small enough to suppress the effect of nonlinear terms in the coupling forces. Thus, for the sake of analytic treatment, we consider a linear chain with linear on-site damping (with damping coefficient $\xi$), subject to the non-degenerate substrate potential ($U_2$). Equations of motion are written in the following general form:

$$\ddot{\varphi}_n + \xi\dot{\varphi}_n + (2\varphi_n - \varphi_{n-1} - \varphi_{n+1}) = -\frac{\partial U_2(\varphi_n)}{\partial \varphi_n} \tag{52}$$

Continuum limit of System (52) is written for the continuous field of displacement $\varphi(x,t)$ as follows:

$$\ddot{\varphi} + \xi\dot{\varphi} - \varphi'' = -\frac{\partial U_2(\varphi)}{\partial \varphi} \tag{53}$$

According to previous treatment (Figure 1 and Appendix A), the potential function obeys the relationships:

$$U_2(\varphi) = \begin{cases} \dfrac{\omega_0^2 \varphi^2}{2} + O(\varphi^3), \varphi \to 0 \\ \dfrac{\omega_0^2 (\varphi - \varphi^*)^2}{2} - Q + O\left[(\varphi - \varphi^*)^3\right], \varphi \to \varphi^* \end{cases} \tag{54}$$

We would like to describe the propagating kink that transmits the system from upper to lower well and suppose the existence of travelling-wave solution in a form

$$\varphi(x,t) = \varphi(x - Vt), \varphi(x \to -\infty) = \varphi^*, \varphi(x \to \infty) = 0 \tag{55}$$

Travelling-wave ansatz (55) converts Equation (53) to an ODE:

$$(V^2 - 1)\varphi_{\zeta\zeta} - V\xi\varphi_\zeta = -\frac{\partial U_2(\varphi)}{\partial \varphi}, \quad \zeta = x - Vt \tag{56}$$

For general potential shape $U_2(\varphi)$ and for nonzero damping coefficient solution of Equation (56) is not known. To obtain a closed-form solution for the transition kink we first adopt piecewise parabolic approximation for the on-site potential:

$$U_2(\varphi) = \begin{cases} \dfrac{\omega_0^2 \varphi^2}{2}, & \varphi < b \\ \dfrac{\omega_0^2 (\varphi - \varphi^*)^2}{2} - Q, & \varphi > b \end{cases} \tag{57}$$

It is obvious that $B = \omega_0^2 b^2 / 2$. Then, to obtain a continuous potential function, one should satisfy the following relationships:

$$\frac{\omega_0^2 b^2}{2} = \frac{\omega_0^2 (b - \varphi^*)^2}{2} - Q \quad \Rightarrow \quad \varphi^* - b = \frac{\sqrt{2(B+Q)}}{\omega_0} \tag{58}$$

So, we see that for selected potential function the parameters are not independent. To derive the expression for the kink, we adopt that the transition between two wells of potential function (57) occurs at the point $\zeta = 0$ $(x = Vt)$. Then, for the region $\zeta > 0$ Equation (56) with potential (57) is reduced to the form:

$$(V^2 - 1)\varphi_{\zeta\zeta} - V\xi\varphi_\zeta = -\omega_0^2 \varphi \tag{59}$$

As it will be demonstrated below, the regime of transitions between the two wells exists only for $V < 1$. The solution of (59) that decays to zero at infinity, is written as:

$$\varphi_+(\zeta) = C_+ \exp(\lambda_+ \zeta), \quad \zeta > 0, \quad \lambda_+ = -\frac{\sqrt{\xi^2 V^2 + 4\omega_0^2 (1-V^2)} + \xi V}{2(1-V^2)} \tag{60}$$

For the region $\xi < 0$ Equation (56) with potential (57) is reduced to the form:

$$(V^2 - 1)\varphi_{\zeta\zeta} - V\xi\varphi_\zeta = -\omega_0^2 (\varphi - \varphi^*) \tag{61}$$

Solution of (61) that satisfies the boundary condition at $\zeta \to -\infty$ is written as:

$$\varphi_-(\zeta) = \varphi^* - C_- \exp(\lambda_- \zeta), \quad \zeta < 0, \quad \lambda_- = \frac{\sqrt{\xi^2 V^2 + 4\omega_0^2 (1-V^2)} - \xi V}{2(1-V^2)} \tag{62}$$

Expressions (60) and (62) should satisfy the following matching conditions at $\zeta = 0$:

$$\varphi_+(0) = \varphi_-(0) = b, \quad \left.\frac{d\varphi_+(\zeta)}{d\zeta}\right|_{\zeta=0} = \left.\frac{d\varphi_-(\zeta)}{d\zeta}\right|_{\zeta=0} \quad (63)$$

Matching conditions (63) yield the following expressions:

$$C_+ = b, \quad C_- = \varphi^* - b, \quad C_+ \lambda_+ = -C_- \lambda_- \quad (64)$$

Substitution of (58), (60) and (62) into (64) yields:

$$\frac{\sqrt{\xi^2 V^2 + 4\omega_0^2 (1-V^2)} - \xi V}{\sqrt{\xi^2 V^2 + 4\omega_0^2 (1-V^2)} + \xi V} = \sqrt{\frac{B}{B+Q}} \quad (65)$$

It is easy to see that for $\xi = 0$ Equation (65) is automatically satisfied for any velocity $V$, but only for the degenerate case $Q = 0$. In the non-degenerate case and for nonzero damping Equation (65) determines unique velocity of the kink. The front velocity can be expressed explicitly as follows:

$$V = \frac{\omega_0 (1-a)}{\sqrt{a\xi^2 + \omega_0^2 (1-a)^2}} \leq 1 \quad a = \sqrt{\frac{B}{B+Q}} \quad (66)$$

The results of front velocity as a function of damping per (66) compared to results obtained from numerical simulations are plotted in Figure 20. Each curve corresponds to a different value of nonlinear coefficient $\beta$. At very high damping we observe that the curves tend to the analytical proposition. As expected, at the lower damping (under $\xi < 0.1$) the continuum model fails to describe the dynamics, as the effect of cubic nonlinearity becomes dominant and the front becomes accompanied by considerable oscillations within the tail region responsible for the radiative damping.

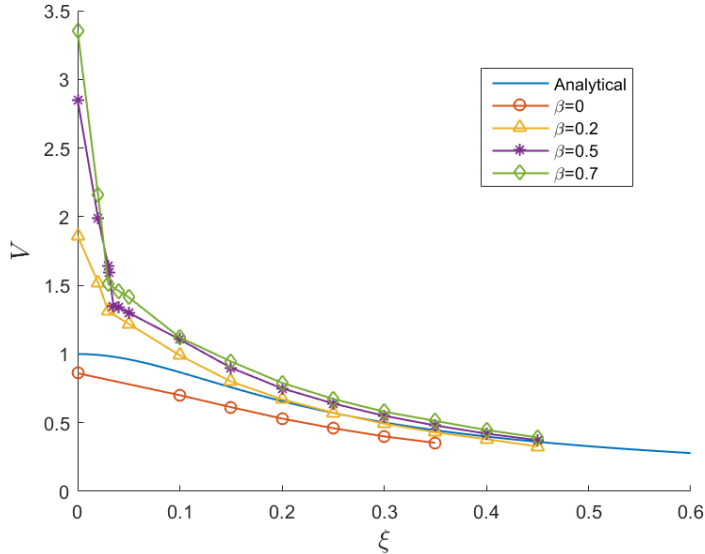

*Figure 20 – Front velocity as a function of damping coefficient, solid blue – analytical, 'o' - beta=0, 'triangles' – $\beta = 0.2$, '*' - $\beta = 0.5$, 'diamonds' – $\beta = 0.7$ ; gradient potential: cubic; on-site potential: bi-parabolic with $Q = 0.5, B = 0.5, \omega_0 = 0.5$. Initial conditions: Impulse 10.*

.

**Concluding remarks**

It seems that the most important finding of the current work is the general character of the described pattern of the front propagation - high velocity, narrow front, low wavenumber of the oscillatory tail, extreme energy concentration in the front zone. These peculiarities make such propagating fronts qualitatively similar to the shock waves. The aforementioned pattern reveals itself for broad range of the coupling potentials (we considered the FPU and LJ models) and is relatively not sensitive to particular shape of the on-site potential. Properties of the front can be described in the framework of the reduced analytic model with single a degree of freedom and appropriate boundary conditions. It was demonstrated that account of the next-nearest-neighbor interactions, the external pre-load and the weak on-site linear damping can be treated as perturbations of the reduced model. Besides, one observes that the external pre-load can be used to modify the front propagation velocity. The case of high damping turns out even easier – this asymptotic limit corresponds to a simple continuous problem.

The treatment presented above leaves many questions for further investigation. First, the derived simplified model works fine only if the front propagation velocity is relatively high. This regime requires considerable energy release at every site. The case of lower energetic effects requires further exploration; one may expect that the role of the gradient nonlinearity in this situation will be less significant. Other interesting problem is possible extension of the simplified local model for higher dimensions.

**Acknowledgment**

The authors are very grateful to Israel Science Foundation (grant 838/13) for financial support

# Appendix

The potentials in Figure 1 are characterized by 3 quantities $Q, B, \varphi^*$. We bring three realizations that conform to these characteristics:

1) Bi-parabolic potential with same curvatures of the two wells $\omega_0$:

$$U_{b-p}(\varphi_n) = \begin{cases} \dfrac{\omega_0^2}{2}\varphi_n^2 & \varphi_n \leq b = \dfrac{\sqrt{2B}}{\omega_0} \\ \dfrac{\omega_0^2}{2}(\varphi_n - \varphi^*)^2 - Q & \varphi_n > b \end{cases} \tag{A1}$$

Here, $\varphi^* = \dfrac{\sqrt{2(Q+B)}}{\omega_0} + \dfrac{\sqrt{2B}}{\omega_0}$

2) 4$^{th}$ order polynomial potential

$$U_4(\varphi_n) = a_2\varphi_n^2 + a_3\varphi_n^3 + a_4\varphi_n^4 \tag{A2}$$

3) 6$^{th}$ order polynomial:

$$U_6(\varphi_n) = b_2\varphi_n^2 + b_3\varphi_n^3 + b_4\varphi_n^4 + b_5\varphi_n^5 + b_6\varphi_n^6 \tag{A3}$$

For the 4$^{th}$ order polynomial, the constraints on $B, Q, \varphi^*$ uniquely define all coefficients. For the 6$^{th}$ order polynomial there is more freedom with 5 coefficients to choose. To obtain the essential deviation from the bi-parabolic potential, the coefficients are chosen to annihilate the second derivative at the maximum (additional condition stems from the fact that the third derivative in this point also must be zero).